\documentclass[12pt]{article}
\usepackage{amsfonts, euscript}

\begin{document}
\rightline{RUB-TP2-15/02}

\vspace{0.3cm}
\begin{center}
{\Large Genuine twist-3 contributions to the generalized parton distributions from instantons}\\

\vspace{0.4cm}

{D. V. Kiptily$^a$, M. V. Polyakov$^{a,b}$}

\vspace{0.35cm}
$^a$Institut f\"ur Theoretische Physik II,
Ruhr--Universit\"at Bochum, D--44780 Bochum, Germany\\
$^b$Petersburg Nuclear Physics
Institute, Gatchina, St.\ Petersburg 188350, Russia
\end{center}

\begin{abstract}
The genuine twist-3 quark-gluon ($\bar q G q$) contributions to
the Generalized Parton Distributions (GPDs) are estimated in the
model of the instanton vacuum. These twist-3 effects are found to
be parametrically suppressed relative to the ``kinematical''
twist-3 ones due to the small packing fraction of the instanton
vacuum. We derived exact sum rules for the twist-3 GPDs.
\end{abstract}

\section{Twist-3 DVCS amplitude}

Deeply Virtual Compton Scattering (DVCS) \cite{Mul94, Ji97b} is a subject of rather
intensive investigations during the last few years
(for a detailed recent review of DVCS see \cite{Bel01}).
It is a two-photon process:
\begin{equation}
\gamma^*(q) + N(P) \to \gamma'(q') + N'(P')\, ,
\end{equation}
in which the initial virtual photon $\gamma^*$, being produced by a lepton,
interacts with a parton of the initial nucleon $N$, the final real photon $\gamma'$
is radiated by the parton, and the nucleon $N'$ in the final state is formed
by the active parton and the rest of the initial nucleon $N$.
The reaction is considered in the Bjorken limit, when the virtuality
of the initial photon $Q^2=-q^2 \gg P^2$ is large, Bjorken variable
$x_{\mbox{\scriptsize Bj}}=Q^2/2(P \cdot q)$ is fixed
and the momentum transfer squared is small: $(P'-P)^2 \ll Q^2$.
The first experimental results on DVCS can be found in Refs.~\cite{H1,Amarian,JLAB,ChAsy}.

We will work in the infinite momentum frame defined by the light-cone vectors
$n^\mu = 1 / (\bar P_+ \sqrt{2}) (1,0,0,-1)$ and
$p^\mu = \bar P_+ / \sqrt{2} (1,0,0,1)$, with $\bar P^\mu = (P + P')^\mu /2$
being the average hadron momentum
(note, that $ n^2 = p^2 = 0 $ and $ n \cdot p = n \cdot \bar P = 1 $).
Let us define the momentum transfer as $\Delta^\mu = (P'-P)^\mu$
and its square as $t=(P'-P)^2$.
In this paper we shall use a kinematical variable $\xi$,
analogous to the Bjorken one, defined as $\xi = - (n \cdot \Delta) / 2$.
A more detailed description of the DVCS kinematics can be found, e.g., in \cite{Bel01}.

To leading order in the electromagnetic coupling constant,
the amplitude of DVCS is proportional to the hadronic tensor $H_{\mu \nu}$,
\begin{equation}
H_{\mu \nu} = - i \int d^4x\, e^{-iqx}
                  \left< P'| T J_\mu(x) J_\nu(0) |P \right>\, .
\end{equation}
To order $(1/Q)^0$, it can be expressed through the non-diagonal hadron matrix elements
${\cal F}_\alpha$ and $\widetilde {\cal F}_\alpha$ of gauge-invariant light-cone bilocal operators:
\begin{eqnarray}
H_{\mu \nu} &=& \frac12 \left( - g^\perp_{\mu \nu} \right)
                        \int^1_{-1} dx\, C_+(x, \xi) \; n^\alpha {\cal F}_\alpha (x, \xi, t)
\nonumber
\\
&&
            {} -   \frac{i}{2} \varepsilon^\perp_{\mu \nu}
                    \int^1_{-1} dx\, C_-(x, \xi) \; n^\alpha \widetilde {\cal F}_\alpha (x, \xi, t)
           + {\cal O} \left( \frac{1}{Q} \right)\, ,
\end{eqnarray}
where ${\cal F}_\alpha$ and $\widetilde {\cal F}_\alpha$ are defined as
\begin{eqnarray}
{\cal F}_\alpha (x, \xi, t)
&=&
\int\limits_{-\infty}^{+\infty} \frac{d\lambda}{2\pi} \, e^{- i \lambda x}
       \left<P',S'\left|
          \bar \psi \left(\frac{\lambda}{2}n \right) \gamma_\alpha
          \psi \left( -\frac{\lambda}{2}n \right)
       \right|P,S\right>\, ,
\\
\widetilde {\cal F}_\alpha (x, \xi, t)
&=&
\int\limits_{-\infty}^{+\infty} \frac{d\lambda}{2\pi} \, e^{- i \lambda x}
       \left<P',S'\left|
          \bar \psi \left(\frac{\lambda}{2}n \right) \gamma_\alpha \gamma_5
          \psi \left( -\frac{\lambda}{2}n \right)
       \right|P,S\right>
\end{eqnarray}
(let us note, that here we do not write explicitly the phase factors
which make the bilocal operators to be those gauge-invariant).
The coefficient functions $C_\pm (x, \xi)$ are
\begin{equation}
C_\pm (x, \xi)  =   \frac1{x - \xi + i \varepsilon}
               \pm  \frac1{x + \xi - i \varepsilon}
\, ,
\end{equation}
and the transverse tensors $- g^\perp_{\mu \nu}$ and $\varepsilon^\perp_{\mu \nu}$ are
\begin{equation}
- g^\perp_{\mu \nu} = - g_{\mu \nu} + n_\mu p_\nu + n_\nu p_\mu\, ,
\qquad
\varepsilon^\perp_{\mu \nu} = \varepsilon_{\mu \nu \lambda \sigma} n^\lambda p^\sigma\, .
\end{equation}
In general, the functions ${\cal F}_\alpha$ and $\widetilde {\cal F}_\alpha$
contain both the purely quark and mixed quark-gluon contributions.
However, in the considered case,
the quark-gluon contribution vanishes due to contraction with the light-cone vector $n^\alpha$.
Therefore the scalar products $n^\alpha {\cal F}_\alpha$ and $n^\alpha \widetilde {\cal F}_\alpha$
can be parameterized in terms of four functions
$H_q(x,\xi,t)$, $E_q(x,\xi,t)$ and $\widetilde H_q(x,\xi,t)$, $\widetilde E_q(x,\xi,t)$
in the following way \cite{Ji97b}:
\begin{eqnarray}
n^\alpha {\cal F}_\alpha &=&
            H_q(x,\xi,t)\; \bar N(P',S') \hat n N(P,S)
          + E_q(x,\xi,t)\; \bar N(P',S')
                              \frac{i \sigma^{\mu\nu} n_\mu \Delta_\nu}{2M}
                           N(P,S)\, ,
\label{twist2}
\\
n^\alpha \widetilde {\cal F}_\alpha &=&
            \widetilde H_q(x,\xi,t)\; \bar N(P',S') \hat n \gamma_5 N(P,S)
          + \widetilde E_q(x,\xi,t)\; \bar N(P',S')
                                     \frac{\gamma_5 \Delta \cdot n}{2M}
                                  N(P,S)\, .
\end{eqnarray}
Here $N(P,S)$ is the Dirac bispinor of a hadron (a proton in our case),
$M$ is the hadron mass and we use the notation $\hat x = x^\mu \gamma_\mu$ throughout the paper.
The functions $H$, $E$ and $\widetilde H$, $\widetilde E$
are the  Generalized Parton Distributions (GPD) (for a review see Refs. \cite{Ji98, Rad01a, Go01}).
In the forward limit $P'=P$ the distributions $H_q$ and $\widetilde H_q$  coincide
with the usual parton densities $q(x)$ and $\Delta q(x)$.
The distributions $E_q$ and $\widetilde E_q$ are completely new ones.
Thus the investigation of DVCS allows us to get an additional information
about the structure of hadrons. In particular, the GPDs contain an information
about the hadron spin distribution among the partons of the hadron \cite{Ji97b, Ji97a}.

The amplitude of DVCS has corrections of two types: firstly, loop corrections
in powers of the electromagnetic and strong coupling constants
and, secondly, higher twist corrections in powers of $1/Q$.
We will not consider the loop corrections  in this paper.
Making allowance for the first correction in powers of $1/Q$,
the hadronic tensor $H_{\mu \nu}$ takes the following form
 \cite{Pen00, Bel00}
(for the pion target see \cite{Ani01,Rad01b}):
\begin{eqnarray}
H_{\mu \nu} &=& \frac12 \left[
                             \left( - g^\perp_{\mu \nu} \right)
                           - \frac{\Delta^\perp_\mu \bar P_\nu}{\left( \bar P \cdot q' \right)}
                        \right]
                        \int^1_{-1} dx\, C_+(x, \xi) \; n^\alpha {\cal F}_\alpha (x, \xi, t)
\nonumber
\\
            &+& \frac{i}{2} \varepsilon^{\perp \lambda}_\mu
                        \left[
                                 \left( - g^\perp_{\lambda \nu} \right)
                               - \frac{\Delta^\perp_\lambda \bar P_\nu}{\left( \bar P \cdot q' \right)}
                            \right]
                        \int^1_{-1} dx\, C_-(x, \xi) \; n^\alpha \widetilde {\cal F}_\alpha (x, \xi, t)
\nonumber
\\
            &-& \frac{(q + 4 \xi \bar P)_\mu}{2 \left( \bar P \cdot q' \right)}
            \left[
                     \left( - g^\perp_{\lambda \nu} \right)
                   - \frac{\Delta^\perp_\lambda \bar P_\nu}{\left( \bar P \cdot q' \right)}
                \right]
        \int^1_{-1} dx\,
        \left\{
             {\cal F}^\lambda_\perp (x, \xi, t) \; C_+(x, \xi)
        \right.
\nonumber
\\
&&              \qquad\qquad\qquad\qquad\qquad\qquad\qquad\qquad\qquad
                \left.
                   {}
           - i \varepsilon^{\lambda \sigma}_\perp \widetilde {\cal F}^\perp_\sigma (x, \xi, t) \; C_-(x, \xi)
        \right\}
\nonumber
\\
            &+& {\cal O} \left( \frac{1}{Q^2} \right) .
\end{eqnarray}
Note that $H_{\mu \nu}$ satisfies the transversity conditions
\begin{eqnarray}
q^\mu H_{\mu \nu}  = 0 ,\quad
H_{\mu \nu} (q')^\nu = 0.
\end{eqnarray}
We see that the correction contains a term proportional to the transverse part
of the vector functions ${\cal F}_\alpha$ and $\widetilde {\cal F}_\alpha$.
This means that, in comparison to the leading order, to order $1/Q$ we deal with corrections
of two types: those ``kinematical'' associated with purely quark operators
and ``genuine twist-3'' ones originated from the quark-gluon operators, which have the form
$\bar\psi G \psi$, where $\psi$ is the operator of the quark field and $G$ denotes
the operator of the gluon field strength tensor. The ``kinematical'' contribution
can be expressed through the twist-2 quark GPDs
$H$, $E$, $\widetilde H$, $\widetilde E$ \cite{Bel00,Ani01,Rad01b,Kiv01a,Kiv00}.

In the case of polarized DIS the quark-gluon correction is supposed to be small
relative to the ``kinematical'' one. This assumption is known as
Wadzura-Wilczek approximation \cite{Wan77}. The experimental data \cite{Ant02, Mit99, Bos00}
point at its reliability.
It is assumed the same approximation to be valid in the case of DVCS.
The aim of this paper is to test this hypothesis in the model of the instanton vacuum \cite{DI84&86}.

A method based on the model of the instanton vacuum, which allows
to calculate hadronic matrix elements of the quark-gluon operators, was suggested \cite{DI96}.
In the case of DIS it was shown \cite{Bal97, Dre00} that the operators of twist-3 are
of order $(\rho/R)^4\ln(\rho/R)$,
where $\rho$ is an average size of instantons and $R$ is an average distance
between instantons in the instanton medium. Thus, the matrix elements are  parametrically
small due to the packing fraction $\rho/R \sim 1/3$. In this paper we apply this analysis
to the nondiagonal hadronic matrix elements of the quark-gluon operators
in the framework of DVCS. We show that in the model of the instanton vacuum they are
parametrically suppressed by the packing fraction of the instanton vacuum. In the forward limit
we reproduce the results for the diagonal case, that is for polarized DIS.

\section{``Genuine twist-3'' effects}

Let us consider the following vector and axial vector bilocal operators
\begin{eqnarray}
\label{V}
V_\alpha (x,-x) &=& \bar \psi(x) \gamma_\alpha [x,-x] \psi(-x)\, ,
\\
\label{PV}
A_\alpha (x,-x) &=& \bar \psi(x) \gamma_\alpha \gamma_5 [x,-x] \psi(-x)\, .
\end{eqnarray}
Here
\begin{equation}
[x,y]=P \exp \left\{
               ig \int_0^1 du \, (x-y)^\nu A_\nu (ux+(1-u)y)
             \right\}
\end{equation}
is the phase factor ordered along a straight line connecting the points $x$ and $y$.
In what follows we will not write it explicitly assuming its presence in all non-local gauge invariant operators.
In oder to separate the ``genuine twist-3'' contributions to the operators (\ref{V}), (\ref{PV}) from
the ``kinematical'' ones we shall use a technique elaborated in \cite{Bel00,Ani01,Rad01b,Kiv01a,Kiv00}.
According to the result obtained by Balitsky and Braun \cite{Bal83, Bal88} in the case
of massless fermion operators there are the following operator identities
($\varepsilon_{0123} = + 1$):
\begin{eqnarray}
\bar \psi(x) \gamma_\alpha \psi(-x) &=&
                \frac{\partial}{\partial x^\alpha}
                \int_0^1 du \, \bar \psi(ux) \hat x \psi(-ux)
\nonumber
\\
          &-& i \varepsilon_{\alpha \beta \lambda \sigma}
                \int_0^1 du \, u x^\beta \mathfrak{D}^\lambda
                \left[
                   \bar \psi(ux) \gamma^\sigma \gamma_5 \psi(-ux)
                \right]
           +    \int_0^1 du \, \mathfrak{G}_\alpha(ux)\, ,
\label{decV1}
\\
\bar \psi(x) \gamma_\alpha \gamma_5 \psi(-x) &=&
                \frac{\partial}{\partial x^\alpha}
                \int_0^1 du\, \bar \psi(ux) \hat x \gamma_5 \psi(-ux)
\nonumber
\\
          &-& i \varepsilon_{\alpha \beta \lambda \sigma}
                \int_0^1 du\, u x^\beta \mathfrak{D}^\lambda
                \left[
                   \bar \psi(ux) \gamma^\sigma \psi(-ux)
                \right]
           +    \int_0^1 du\, \mathfrak{G}_{5 \alpha}(ux)\, ,
\label{decPV1}
\end{eqnarray}
where
\begin{eqnarray}
\mathfrak{G}_\alpha(x) &=& g \int_{-1}^1 dv \,
\bar\psi(x) \left[
               \varepsilon_{\alpha\mu\nu\tau} G^{\mu\rho}(vx) x_\rho x^\nu \gamma^\tau \gamma_5
             - i v G_{\alpha\rho}(vx) x^\rho \hat x
            \right] \psi(-x)\, ,
\label{Gterm}
\\
\mathfrak{G}_{5 \alpha}(x) &=& g \int_{-1}^1 dv\,
\bar\psi(x) \left[
               \varepsilon_{\alpha\mu\nu\tau} G^{\mu\rho}(vx) x_\rho x^\nu \gamma^\tau \gamma_5
             - i v G_{\alpha\rho}(vx) x^\rho \hat x
            \right] \gamma_5 \psi(-x)\, .
\label{G5term}
\end{eqnarray}
In these equations the derivative $\mathfrak{D}^\lambda$ is defined for any combination $\Gamma$
of Dirac $\gamma$-matrices as
\begin{equation}
\label{Dder}
\mathfrak{D}^\lambda
   \left\{
      \bar \psi(x) \Gamma [x,-x] \psi(-x)
   \right\} = \lim_{y \to 0} \frac{\partial}{\partial y_\lambda}
                \left\{
           \bar \psi(x+y) \Gamma [x+y,-x+y] \psi(-x+y)
        \right\}.
\end{equation}
Later we will use the fact that the non-diagonal matrix element of (\ref{Dder})  is proportional
to the momentum transfer:
\begin{equation}
\label{Dprop}
\left< P',S' \left|
   \mathfrak{D}^\lambda
      \left\{
         \bar \psi(x) \Gamma [x,-x] \psi(-x)
      \right\}
\right| P,S \right> = i(P'-P)^\lambda
                                 \left< P',S' \left|
                    \bar \psi(x) \Gamma [x,-x] \psi(-x)
                 \right| P,S \right>.
\end{equation}
If we took the fermion mass into account, there would appear an additional mass term
in the expression (\ref{decPV1}) for the axial vector bilocal operator
\begin{equation}
\label{mterm}
{} + 2 i m \int_0^1 du\, u \;
       \bar \psi(ux) i \sigma_{\alpha \beta} x^\beta \gamma_5 \psi(-ux)  .
\end{equation}

We would now like to express the operators (\ref{V}), (\ref{PV}) through the symmetrical ones:
\begin{eqnarray}
V_{sym} (x,-x) &=& \bar \psi(x) \hat x [x,-x] \psi(-x)\, ,
\\
A_{sym} (x,-x) &=& \bar \psi(x) \hat x \gamma_5 [x,-x] \psi(-x)\, .
\end{eqnarray}
These operators are ``symmetrical'' in a sense that on light cone ($x^2 = 0$) they expand in series
of symmetrical traceless local operators. In order to do so we should solve
the system of ordinary differential equations (\ref{decV1}), (\ref{decPV1})
with respect to $V_\alpha (x,-x)$ and $A_\alpha (x,-x)$ (see Appendix of ref. \cite{Kiv00}
for details). The vector solution can be presented in the following form:
\begin{equation}
V_{\alpha}(x,-x) = V_{\alpha}^{WW}(x,-x) + V_{\alpha}^{tw3}(x,-x)\, ,
\label{decV2}
\end{equation}
where
$$
V_{\alpha}^{WW}(x,-x) =
          \frac12 \int_0^1 du\, \left\{
                               e^{(1-u)[(x \cdot \mathfrak{D})^2-x^2{\mathfrak{D}}^2]^{1/2}} +
                   e^{-(1-u)[(x \cdot \mathfrak{D})^2-x^2{\mathfrak{D}}^2]^{1/2}}
                            \right\}
$$
$$
      \left\{
            \left[
              x_\alpha {\mathfrak{D}}^2 - \left( x \cdot \mathfrak{D} \right) {\mathfrak{D}}_\alpha
            \right]
               \int_0^u dv\, v\; \bar \psi(vx) \hat x  [vx,-vx] \psi(-vx)
         +  \frac{\partial}{\partial x^\alpha}
                  \left[
             \bar \psi(ux) \hat x  [ux,-ux] \psi(-ux)
          \right]
      \right.
$$
\begin{equation}
          \left.
      {} - i \varepsilon_{\alpha \beta \lambda \sigma} x^\beta \mathfrak{D}^\lambda
             \frac{\partial}{\partial x_\sigma}
            \int_0^u dv\; \bar \psi(vx) \hat x \gamma_5 [vx,-vx] \psi(-vx)
      \right\}   ,
\label{decVWW}
\end{equation}
$$
V_{\alpha}^{tw3}(x,-x) =
          \frac12 \int_0^1 du\, \left\{
                               e^{(1-u)[(x \cdot \mathfrak{D})^2-x^2{\mathfrak{D}}^2]^{1/2}} +
                   e^{-(1-u)[(x \cdot \mathfrak{D})^2-x^2{\mathfrak{D}}^2]^{1/2}}
                            \right\}
$$
$$
          \left\{
          {} - i \left[
                x^2 \mathfrak{D}_\alpha - x_\alpha \left( x \cdot \mathfrak{D} \right)
             \right]
           \int_0^u dv\, v \int_{-v}^v dt\;
              \bar \psi(vx) [vx,tx] g G_{\mu\nu}(tx) x^\mu \gamma^\nu  [tx,-vx] \psi(-vx)
      \right.
$$
\begin{equation}
      \left.
         {} + \mathfrak{G}_\alpha(ux)
        - i \varepsilon_{\alpha \beta \lambda \sigma} x^\beta \mathfrak{D}^\lambda
                          \int_0^u dv\, \mathfrak{G}_5^\sigma(vx)
      \right\} .
\label{decVtw3}
\end{equation}
Here $V_{\alpha}^{WW}$ is the ``kinematical'' (or Wandzura-Wilczek) contribution
to the bilocal vector operator (\ref{V}).
The WW-contribution for the spin-$\frac12$ target was discussed in details in Refs.~\cite{Bel00,Kiv01a}.
The second term $V_{\alpha}^{tw3}$ in (\ref{decV2}) is the ``genuine twist-3''
(quark-gluon) contribution to the vector operator.
In the case $x^2=0$ the formulae (\ref{decV2})-(\ref{decVtw3})
reproduce the result of Belitsky and M\"uller~\cite{Bel00}. It is of necessity to note that
the ``genuine twist-3'' contribution to the scalar product $x^{\alpha} V_{\alpha}(x,-x)$
vanishes, with the first term of the ``kinematical'' contribution only being survived.
In the case when $x^\alpha = \lambda n^\alpha / 2$ this means that the quark-gluon contribution
to the function $ n^\alpha \cal{F}_\alpha $ vanishes, giving rise to
the decomposition (\ref{twist2}).

The corresponding expressions for the axial vector operator
(\ref{PV}) can be obtained by the following substitution in Eqs. (\ref{decV2})-(\ref{decVtw3}):
\begin{eqnarray*}
\Gamma \psi  & \longrightarrow &   \Gamma \gamma_5 \psi  \, ,
\\
\mathfrak{G}_\alpha  & \longleftrightarrow &   \mathfrak{G}_{5 \alpha}  \, .
\end{eqnarray*}

\section{Diagonal nucleon matrix elements}

Let us consider polarized structure functions defined as integrals over
diagonal matrix elements of string operators (for a review see \cite{Ans95,Jaf96}
and references therein):
\begin{eqnarray}
\frac1{2M}
        \int\limits_{-\infty}^{+\infty} \frac{d\lambda}{2\pi} \, e^{- i \lambda x}
    \left< P,S \left|
           \bar \psi \left(
                    \frac{\lambda n}{2}
             \right)
       \gamma_\mu \gamma_5
       \psi \left(
               - \frac{\lambda n}{2}
        \right)
    \right| P,S \right> &=& g_1(x) (S \cdot n) p_\mu + g_T(x) S_{\perp \mu}
\nonumber
\\
                        &+& g_3(x) M^2 (S \cdot n) n_\mu \,  ,
\end{eqnarray}
\begin{eqnarray}
\frac1{2M}
        \int\limits_{-\infty}^{+\infty} \frac{d\lambda}{2\pi} \, e^{- i \lambda x}
    \left< P,S \left|
           \bar \psi \left(
                    \frac{\lambda n}{2}
             \right)
       i \sigma_{\mu \nu} \gamma_5
       \psi \left(
               - \frac{\lambda n}{2}
        \right)
    \right| P,S \right> &=& h_1(x) \frac1M (S_{\perp \mu} p_\nu - S_{\perp \nu} p_\mu)
\nonumber
\\
                        &+& h_L(x) M (S \cdot n) (p_\mu n_\nu - p_\nu n_\mu)
\nonumber
\\
                        &+& h_3(x) M (S_{\perp \mu} n_\nu - S_{\perp \nu} n_\mu),
\end{eqnarray}
where the spin 4-vector is normalized as $S^2 = - 1$.
Taking the matrix element of (\ref{decPV1}) with the mass term (\ref{mterm})
we arrive at formulas known as Wandzura-Wilczek relations (\cite{Wan77,Tan94}).
For the region $0<x<1$ we have:
$$
S_{\perp \alpha} \left\{
                    g_T(x)
              - \int_x^1 \frac{dy}{y}\, g_1(y)
              - \frac{m}{M}
                   \left[
                  \frac{h_1(x)}{x} - \int_x^1 \frac{dy}{y^2}\, h_1(y)
               \right]
                 \right\}  =
$$
\begin{equation}
{} = \frac1{2M}\,
               \int_x^1 \frac{dy}{y}\,
               \int\limits_{-\infty}^{+\infty} \frac{d\lambda}{2\pi} \, e^{- i \lambda y}
              \left< P,S_\perp \left|
             \mathfrak{G}_{5 \alpha}
                \left(
                       \frac{\lambda n}{2}
                \right)
              \right| P,S_\perp \right>  .
\end{equation}
An analogous equation for the region $-1<x<0$ reads:
$$
S_{\perp \alpha} \left\{
                    g_T(x)
              + \int_{-1}^x \frac{dy}{y}\, g_1(y)
              - \frac{m}{M}
                   \left[
                  \frac{h_1(x)}{x} + \int_{-1}^x \frac{dy}{y^2}\, h_1(y)
               \right]
                 \right\}  =
$$
\begin{equation}
{} = - \frac1{2M}\,
                  \int_{-1}^x \frac{dy}{y}\,
                  \int\limits_{-\infty}^{+\infty} \frac{d\lambda}{2\pi} \, e^{- i \lambda y}
                 \left< P,S_\perp \left|
                \mathfrak{G}_{5 \alpha}
                   \left(
                          \frac{\lambda n}{2}
                   \right)
                 \right| P,S_\perp \right>  .
\end{equation}
It is easy to get the Wanzsura-Wilczek relations for the Mellin moments.
For example, in the massless case they have the form:
\begin{equation}
S_{\perp \alpha} \int_{-1}^1 dx\, x^m \left[
                                      g_T(x) - \frac1{m+1} g_1(x)
                                    \right] =
\frac{(-i)^m}{2M(m+1)}
   \left.
      \frac{\partial^m}{\partial \lambda^m}
      \left< P,S_\perp \left|
     \mathfrak{G}_{5 \alpha}
        \left(
               \frac{\lambda n}{2}
        \right)
      \right| P,S_\perp \right>
   \right|_{\lambda=0}  .
\label{WWmom}
\end{equation}
In particular, for $m=0$ we have:
\begin{equation}
\int_{-1}^1 dx\, \left[ g_T(x) - g_1(x) \right] = 0  ;
\label{WW0mom}
\end{equation}
for $m=1$ we get:
\begin{equation}
\int_{-1}^1 dx\, x \left[ g_T(x) - \frac12 g_1(x) \right] = 0  ;
\label{WW1mom}
\end{equation}
and for $m=2$ we arrive at:
$$
S_{\perp \alpha} \int_{-1}^1 dx\, x^2 \left[
                                         g_T(x) - \frac13 g_1(x)
                                      \right] =
$$
\begin{equation}
{} = \frac1{6M}
             \left< P,S_\perp \left| g\,
                         \bar\psi
                            \left[
                               \widetilde G_{\alpha \beta} n^\beta \hat n
                       - n_{\alpha} \widetilde G_{\lambda \beta} n^\beta \gamma^\lambda
                        \right] \psi
             \right| P,S_\perp \right>   .
\label{WW2mom}
\end{equation}
The second term in (\ref{WW2mom}) is purely longitudinal in
$\alpha$ and does not give any contribution, therefore the
transverse part of (\ref{WW2mom}) takes the form
\begin{equation}
\label{WW2momperp}
S_{\perp \alpha} \int_{-1}^1 dx\, x^2 \left[
                                      g_T(x) - \frac13 g_1(x)
                                    \right]
= \frac1{6M}
          \left< P,S_\perp \left|
             g\, \bar\psi  \widetilde G_{\alpha \beta}  n^\beta \hat n \psi
          \right| P,S_\perp \right>, \qquad \alpha = 1,2   .
\end{equation}

The validity of (\ref{WW0mom})-(\ref{WW2mom}) follows from the observation that
$\mathfrak{G}_5(\lambda n / 2) \sim \lambda^2$ and, consequently, the first
non-vanishing moment in (\ref{WWmom}) is the $x^2$-moment.
The eqs. (\ref{WW0mom}), (\ref{WW1mom}) known as Burkhardt-Cottingham \cite{Bur70} and
Efremov-Leader-Teryaev \cite{Efr97} sum rules correspondingly.
Using the definition of the constant $d^{(2)}$, which measures the effect of genuine
twist-3 contributions:
\begin{equation}
\label{d2}
\int_{-1}^1 dx\, x^2 \left[ g_T(x) - \frac13 g_1(x) \right] = \frac23 \, d^{(2)}  ,
\end{equation}
we come to the identity
\begin{equation}
\label{MatrEl=d2}
S_{\perp}^{\alpha} \left< P,S_\perp \left|
                      g\, \bar\psi  \widetilde G_{\alpha \beta}  n^\beta \hat n \psi
                   \right| P,S_\perp \right>
           = - 4 M d^{(2)}  ,
\end{equation}
which illustrates that the constant $d^{(2)}$ characterizes the correlations of quarks
and gluons in a hadron.

\section{Nondiagonal nucleon matrix elements}

In this section we would like to concentrate on gluon part $V_\alpha^{tw3} (x,-x)$
of eqs. (\ref{decV2})-(\ref{decVtw3}) with $x= \frac12 \lambda n$.
Fourier transformation of its nondiagonal nucleon matrix element
\begin{equation}
{\cal F}_\alpha^{tw3}(P',S';P,S; x) =
\int\limits_{-\infty}^{+\infty} \frac{d \lambda}{2 \pi} \, e^{- i \lambda x}
\left< P',S' \left|
   V_\alpha^{tw3} \left( \frac{\lambda n}{2}, - \frac{\lambda n}{2} \right)
\right| P,S \right>
\end{equation}
is the gluon part of the vector GPD, where the spin 4-vectors are normalized as
$S^2 = S'^2 = -1$. Mellin moments of ${\cal F}_\alpha^{tw3}$ are given by
\begin{equation}
\int_{-1}^1 dx\, x^m {\cal F}_\alpha^{tw3}(P',S'; P,S; x) =
       (-i)^m \left.
                 \frac{\partial^m}{\partial \lambda^m}
             \left< P',S' \left|
                    V_\alpha^{tw3} \left( \frac{\lambda n}{2}, - \frac{\lambda n}{2} \right)
                 \right| P,S \right>
          \right|_{\lambda = 0}   .
\label{tw3mmom}
\end{equation}

It is not difficult to predict the results of the calculation of the first moments
in (\ref{tw3mmom}). The $\mathfrak{G}$-term in (\ref{decVtw3}) is proportional to $x^2$
(\ref{Gterm}) and, consequently, to $\lambda^2$. The first and the third terms
in (\ref{decVtw3}) are proportional to $x^3$ and, consequently, to $\lambda^3$.
This means that $x^0$- and $x^1$-moments should vanish; the $\mathfrak{G}$-term
defines the $x^2$-moment, with the other terms of (\ref{decVtw3}) being given
their contributions to the highest moments only. Thus one can get by direct calculation that
\begin{equation}
\label{tw301mom}
\int_{-1}^1 dx\,     {\cal F}_\alpha^{tw3}(x; P',S'; P,S) =
\int_{-1}^1 dx\, x   {\cal F}_\alpha^{tw3}(x; P',S'; P,S) = 0   ,
\end{equation}
These results  mean that the first two moments of the GPDs do not get
gluon contributions, the WW approximation turns out to be exact for them.

The first nontrivial Mellin moment sensitive to the genuine twist-3 contributions
is the $x^2$-moment. It has the following form:
\begin{eqnarray}
\label{tw32mom}
\int_{-1}^1 dx\, x^2 {\cal F}_\alpha^{tw3}(x; P',S'; P,S) &=&
\frac13 \left< P',S' \left| g\,
                       \bar\psi
                          \left[
                             \widetilde G_{\alpha \beta}  n^\beta \hat n
                          \right.
                     \right.
    \right.
\nonumber
\\
&& \qquad\qquad\quad
        \left.     \left.
                      \left.
                         {}
                         - n_\alpha \widetilde G_{\lambda \beta}  n^\beta \gamma^\lambda
                      \right] \gamma_5 \psi
        \right| P,S \right>   .
\end{eqnarray}
The transverse part of eq. (\ref{tw32mom}) has the form
\begin{equation}
\label{tw32momperp}
\int_{-1}^1 dx\, x^2 {\cal F}_\alpha^{tw3}(x; P',S'; P,S) =
\frac13 \left< P',S' \left|
           g\, \bar\psi  \widetilde G_{\alpha \beta}  n^\beta \hat n \gamma_5 \psi
        \right| P,S \right>, \qquad \alpha=1,2.
\end{equation}
The expressions for the axial moments can be obtained using the substitution:
$\Gamma \psi  \longrightarrow   \Gamma \gamma_5 \psi$.

\section{Estimate of the nucleon matrix elements}

In order to estimate the magnitude of the second moments (\ref{WW2momperp}), (\ref{tw32momperp}),
we should analyze the nondiagonal {\it hadron} matrix elements of the quark-gluon operators:
\begin{eqnarray}
\label{O}
O_\perp &=& g\, \bar\psi  \widetilde G_{\perp \lambda}  n^\lambda \hat n \psi  ,
\\
\label{O5}
O_{5 \perp} &=& g\, \bar\psi  \widetilde G_{\perp \lambda}  n^\lambda \hat n \gamma_5 \psi
\end{eqnarray}
(in this section we will use a notation: $O_\perp \equiv O_\alpha, \quad \alpha = 1,2$).
For this purpose we first consider the nondiagonal {\it quark} matrix elements
of the operators (\ref{O}), (\ref{O5}).
A method of calculation of such matrix elements in the model of the instanton vacuum
was suggested in Refs. \cite{DI96, Bal97}. It was developed in the framework
of the Effective Chiral Theory of the nucleon \cite{DI88}
(see also recent reviews \cite{DI96b, DI01} and the references therein).
The essence of the method is in the following.

In the model of the instanton vacuum \cite{DI84&86}  the fermion partition function in the presence of
$N_+$ instantons and $N_-$ anti-instantons can be represented as a functional integral
over the fermion fields with a fermion effective action $S_{\mbox{\scriptsize eff}}[\bar \psi, \psi]$:
\begin{equation}
Z_{N_{\pm}}^{\mbox{\scriptsize fermions}} = Z_{N_{\pm}}
                         C \int {\cal D} \bar \psi {\cal D} \psi
                    \exp \left\{
                        i S_{\mbox{\scriptsize eff}}[\bar \psi, \psi]
                     \right\}  ,
\end{equation}
where in the case of one quark flavor
\begin{equation}
S_{\mbox{\scriptsize eff}}[\bar \psi, \psi] =
                           \int \frac{d^4k}{(2 \pi)^4} \,
                           \bar \psi(k)
                                          \left[
                         \hat k - M(ik)
                      \right]
                                       \psi(k)
\label{Seff}
\end{equation}
and $Z_{N_{\pm}}$ is the variational partition function of the gluodynamics.
Eq. (\ref{Seff}) points out that a quark propagating in the instanton medium
acquires a dynamical momentum-dependent mass $M(ik)$.
It is this mass that leads to the nonvanishing value of the vacuum condensate
$\langle \bar \psi \psi \rangle$ and, consequently, to the spontaneous chiral
symmetry breaking in QCD \cite{DI96b}. The quark propagator
in the instanton vacuum model can be written in the momentum representation as
\begin{equation}
\hat S(k) = i \frac{\hat k + M(ik)}{k^2 - M^2(ik) + i \varepsilon} .
\end{equation}
The dynamical mass $M(ik)$ has the form:
\begin{equation}
M(ik) = M |F(ik)|^2.
\label{mass}
\end{equation}
Here the mass $M \sim \rho / R^2 \sim 345 $ MeV, where $\rho$ is the average size
of the instantons in the instanton medium and $R$ is the average distance between instantons.
The instanton medium is characterized by its ``packing fraction'' $\rho/R$
with a phenomenological value  $\rho/R \sim 1/3$. Thus, the product $M \rho \sim \rho^2 / R^2$
is parametrically small. The function $F(k)$ in (\ref{mass}) is a form factor proportional
to the Fourier transformation of the wave function of the fermion zero-mode:
\begin{eqnarray}
F(k) &=& {}
      -  t \frac{d}{dt}
         \left[
            I_0(t) K_0(t) - I_1(t) K_1(t)
         \right] ,
\qquad t = \frac12 k \rho
  ,
\\
F(k) &=& 1, \qquad \rho \to 0 ,
\nonumber
\end{eqnarray}
where $I_n(t)$ are the modified Bessel functions and $K_n(t)$ are the modified Hankel functions.

The effective action (\ref{Seff}) can be used in the calculation of the quark matrix elements
of the quark-gluon operators (\ref{O})-(\ref{O5}), the operators being replaced by
the effective ones:
\begin{equation}
\left< p',s' \left| O \right| p,s \right>_{\mbox{\scriptsize inst.vac.}} =
\left< p',s' \left| ``O" \right| p,s \right>_{\mbox{\scriptsize eff}}.
\end{equation}
Here $``O"$ is the {\it fermion} operator, which represents the {\it quark-gluon} operator $O$
in the effective fermion theory. The explicit form of $``O"$ was obtained in \cite{DI96,Pol96}.
The effective action and effective operators being known, it is possible
to calculate the quark matrix elements of interest.

The direct calculation for the case of the operators (\ref{O}), (\ref{O5}) results in
the following expressions:
\begin{eqnarray}
\left< p',s' \left| ``O_\perp" \right| p,s \right>_{\mbox{\scriptsize eff}} &=&
\frac12 M \rho^2 \int \frac{d^4k}{(2\pi)^4}\, G(ik)
\nonumber
\\
&&\bar u(p',s') \left[
                   F^* {\bf(} i(k-p) {\bf )} F(ip)\;  \hat n \gamma_5 \hat S(k-p) \hat\Sigma_\perp(k,n)
                \right.
\nonumber
\\
&&              \left.
              {} - F^*(ip') F {\bf (} i(k-p') {\bf )}\; \hat \Sigma_\perp(k,n) \hat S(k-p') \hat n \gamma_5
                \right] u(p,s)
\nonumber
\\
&& {} + (\mbox{corrections}) ,
\label{Vmelgeneral}
\\
\left< p',s' \left| ``O_{5 \perp}" \right| p,s \right>_{\mbox{\scriptsize eff}} &=&
\frac12 M \rho^2 \int \frac{d^4k}{(2\pi)^4}\, G(ik)
\nonumber
\\
&&\bar u(p',s') \left[
                   F^* {\bf (} i(k-p) {\bf )} F(ip)\;  \hat n \hat S(k-p) \hat \Sigma_\perp(k,n)
                \right.
\nonumber
\\
&&          \left.
               {} - F^*(ip')F {\bf (} i(k-p') {\bf )}\; \hat \Sigma_\perp(k,n) \hat S(k-p') \hat n
                \right] u(p,s)
\nonumber
\\
&& {} + (\mbox{corrections})  .
\label{Amelgeneral}
\end{eqnarray}
Here $G(k)$ is the form factor proportional to the Fourier transformation of the one-instanton
dual field strength
\begin{eqnarray}
G(k) &=& 32 \pi^2
         \left[
            \left(
           \frac12 + \frac{4}{t^2}
        \right) K_0(t) +
        \left(
           \frac{2}{t} + \frac{8}{t^3}
        \right) K_1(t) -
        \frac{8}{t^4}
         \right]  ,
\qquad t = k \rho ,
\\
G(k) &=& {} - 4 \pi^2, \qquad \rho \to 0 ,
\end{eqnarray}
and the following definitions have been used:
$$
\hat\Sigma_\alpha(k,n) = \frac{k_\alpha k^\mu}{k^2} n^\nu \sigma_{\mu \nu}
                       + \frac{k^\nu k^\lambda}{k^2} n_\lambda \sigma_{\alpha \nu}
               - \frac12 n^\nu \sigma_{\alpha \nu}   ,
$$
$$
\sigma_{\mu \nu } = \frac{i}{2} [\gamma_\mu, \gamma_\nu], \qquad
        n^\lambda = \frac{1}{\bar p_+ \sqrt{2}} (1;0,0,-1)  ,
$$
$u(p,s)$ - is the quark Dirac bispinor.

Using equations of motion, the expressions (\ref{Vmelgeneral}), (\ref{Amelgeneral})
can be reduced to the form:
\begin{eqnarray}
\label{Vqmel}
\left< p',s' \left| ``O_\perp" \right| p,s \right>_{\mbox{\scriptsize eff}}
&=& M I(p) \; \bar u(p',s')\, \hat n \gamma_\perp \gamma_5\, u(p,s)  ,
\\
\label{Aqmel}
\left< p',s' \left| ``O_{5 \perp}" \right| p,s \right>_{\mbox{\scriptsize eff}}
&=& - \xi M I(p) \; \bar u(p',s')\, \hat n \gamma_\perp\, u(p,s)  ,
\end{eqnarray}
where $\xi = - \frac12 (p'-p) \cdot n$, and the integral $I(p)$
is defined in leading oder of $\rho^2/R^2$ as
\begin{equation}
I(p) = \rho^2 \frac13 \int \frac{d^4 k}{(2\pi)^4}\,
                     \frac{G(k)F^3(k-p)}{(k-p)^2 + M^2 F^2(k-p)}
             \left[
                {} - 1 + 4 \frac{(k \cdot p)^2}{k^2 p^2}
             \right]  ,
\label{I}
\end{equation}
where the integration is performed in Euclidean space with $p^2 = - M^2$.
In the case of diagonal matrix elements we reproduce (up to a sign)
the results of Ref. \cite{Bal97}:
\begin{eqnarray}
\left< p,s \left| ``O_\perp"     \right| p,s \right>_{\mbox{\scriptsize eff}} &=&
4 M d^{(2)}_{\mbox{\scriptsize quark}} \, s_\perp  ,
\\
\left< p,s \left| ``O_{5 \perp}" \right| p,s \right>_{\mbox{\scriptsize eff}} &=& 0  ,
\end{eqnarray}
where
\begin{equation}
d^{(2)}_{\mbox{\scriptsize quark}} = \frac12  I(p)
\end{equation}
is a quark constant analogous to the hadron constant (\ref{d2}) in Eq. (\ref{MatrEl=d2}).
Thus, the nondiagonal quark matrix elements (\ref{Vqmel}), (\ref{Aqmel})
of the twist-3 quark-gluon operators can be expressed through the quark constant:
\begin{eqnarray}
\label{Vqmeld2}
\left< p',s' \left| ``O_\perp" \right| p,s \right>_{\mbox{\scriptsize eff}}
&=&   2 M d^{(2)}_{\mbox{\scriptsize quark}}     \; \bar u(p',s')\, \hat n \gamma_\perp \gamma_5\, u(p,s)  ,
\\
\label{Aqmeld2}
\left< p',s' \left| ``O_{5 \perp}" \right| p,s \right>_{\mbox{\scriptsize eff}}
&=& - 2 \xi M d^{(2)}_{\mbox{\scriptsize quark}} \; \bar u(p',s')\, \hat n \gamma_\perp\, u(p,s)  .
\end{eqnarray}
In the limit of a small average size of instantons $\rho$ we get the following estimate of
the integral $I(p)$ (\ref{I}):
\begin{equation}
\label{smalleness}
d^{(2)}_{\mbox{\scriptsize quark}}\sim I(p) \sim \rho^2 M^2 \ln \rho^2 M^2  ,
\qquad
\rho \ll R .
\end{equation}
The quantity $M \rho \sim \pi \rho^2 / R^2$ is a small parameter in the effective fermion theory and,
thus, we notice the quark matrix elements of the quark-gluon operators to be parametrically
suppressed as a forth power of the ``packing fraction'' $\rho / R$ of the instanton vacuum.
The numerical calculation \cite{Bal97} of $I(p)$ gives $d^{(2)}_{\mbox{\scriptsize quark}} = 0.011$
at the phenomenological value $M \rho = 0.58$.

The non-zero value of $d^{(2)}_{\mbox{\scriptsize quark}}$ appears only at the order
$\sim \pi^2\rho^4/R^4$. At the present stage of development of our theory
we cannot make precise predictions at this order,
as this would require to go beyond the one-instanton approximation.
Therefore we consider the value of
 $d^{(2)}_{\mbox{\scriptsize quark}} = 0.011$
only as an order of magnitude estimate. Using this order of magnitude estimate
we can predict the order of magnitude of $d^{(2)}$ for the nucleon \cite{Bal97}:
\begin{equation}
d^{(2)}\sim 10^{-3},\ \ \ {\rm with}\ \ \ d^{(2)}_p-d^{(2)}_n\sim 1/N_c, \ \ \ d^{(2)}_p+d^{(2)}_n\sim 1/N_c^2.
\end{equation}
Without the suppression due to the instanton packing fraction a ``natural" value of the nucleon $d^{(2)}$
would be of order $\sim \int_0^1 dx\ x^2\ g_1(x)\sim {\rm few}\times 10^{-2}$. A recent experiment at SLAC \cite{Ant02}
quotes the values $d^{(2)}_p=(3.2\pm 1.7)\cdot 10^{-3}$ and $d^{(2)}_n=(7.9\pm 4.8)\cdot 10^{-3}$
at $Q^2 \approx 5$~GeV$^2$. These data hint on possible ``anomalous suppression" of $d^{(2)}$
as predicted in the model of the instanton vacuum \cite{Bal97}.

Using the results (\ref{Vqmeld2}) and (\ref{Aqmeld2}) for the quark matrix elements we
can make a simple estimate of the corresponding nucleon matrix element at $\Delta^2=0$.
In such a kind of estimate we assume that the nucleon matrix element at $\Delta^2=0$ has the same
structure as in Eqs.~(\ref{Vqmeld2},\ref{Aqmeld2}).
Such an assumption corresponds to the use of a simple quark model to ``translate" the quark
matrix elements to the nucleon ones. More sophisticated estimates can be done with help
of the chiral quark-soliton model, see e.g. Refs.~\cite{Bal97,Lee}. Note, however, that for the very fact of
the suppression of the twist-3 nucleon matrix elements the usage of the simple quark model
is sufficient. In the simple quark model of the nucleon we arrive at:
\begin{eqnarray}
\label{Vnucl}
\left< P',S' \left| ``O_\perp" \right| P,S \right>_{\mbox{\scriptsize eff}}
&=&   2 M_N d^{(2)} \;
          \bar N(P',S')\, \hat n \gamma_\perp \gamma_5\, N(P,S)  ,
\\
\label{Anucl}
\left< P',S' \left| ``O_{5 \perp}" \right| P,S \right>_{\mbox{\scriptsize eff}}
&=& - 2 \xi M_N d^{(2)} \;
              \bar N(P',S')\, \hat n \gamma_\perp\, N(P,S)  ,
\end{eqnarray}
where $N(P,S)$ is the nucleon Dirac bispinor. We come to the conclusion that
the hadron matrix elements of the ``genuine twist-3'' (quark-gluon) opeators
are parametrically suppressed in the effective chiral theory
by the ``packing fraction'' of the instanton vacuum.

\section{GPDs decomposition}

The quark twist-2 (WW) contributions to the nucleon GPD ${\cal F}_\perp$
and $\widetilde {\cal F}_\perp$ have been obtained in Refs. \cite{Bel00,Kiv01a}.
In this section we would like to compare the WW contributions
with the genuine twist-3 ones, calculated in the previous section
(to be precise, we will compare their $x^2$-moments).
Following Ref.~\cite{Pen00} we present the ``transverse" part of the GPDs
in the form\footnote{We slightly changed the notations of Ref.~\cite{Pen00}
and added supplementary functions which were missed in Ref.~\cite{Pen00},
see also \cite{Mukherjee:2002xi}.}:
\begin{eqnarray}
\label{QVdec0}
{\cal F}_{\perp \mu} (x, \xi, \Delta)
&=&
\bar N(P',S') \biggl\{\left( H + E \right)\; \gamma_\mu^\perp
\nonumber
\\
&+&    G_1\;   \frac{\Delta_\mu^\perp}{2M}
     + G_2\;   \gamma_\mu^\perp
     + G_3\;   \Delta_\mu^\perp \hat n
     + G_4\; i \varepsilon_{\mu \nu}^\perp \Delta_\perp^\nu \hat n \gamma_5
\biggr\} N(P,S) ,
\\
\label{QAdec0}
\widetilde {\cal F}_{\perp \mu} (x, \xi, \Delta)
&=&
\bar N(P',S') \biggl\{   \widetilde H\; \gamma_\mu^\perp \gamma_5
                       + \widetilde E\; \frac{\Delta_\mu^\perp}{2M} \gamma_5
\nonumber
\\
&+&     \widetilde G_1\;   \frac{\Delta_\mu^\perp}{2M} \gamma_5
      + \widetilde G_2\;   \gamma_\mu^\perp \gamma_5
      + \widetilde G_3\;   \Delta_\mu^\perp \hat n \gamma_5
      + \widetilde G_4\; i \varepsilon_{\mu \nu}^\perp \Delta_\perp^\nu \hat n
\biggr\} N(P,S) .
\end{eqnarray}
Note that in the above equations all functions depend on $x,\xi$ and $\Delta^2$.
The first three moments of the quark contributions have the following structure.

%
%X^0 MOMENTS.
%

The $x^0$-moments are:
\begin{eqnarray}
\int_{-1}^{1} dx\, {\cal F}_{\perp \mu}(x, \xi, \Delta)
&=&
\bar N(P',S') \biggl\{
     G_M(\Delta^2)\; \left[ \gamma_\mu^\perp \right]
\biggr\} N(P,S),
\label{V0mom}
\\
\int_{-1}^{1} dx\, \widetilde {\cal F}_{\perp \mu}(x, \xi, \Delta)
&=&
\bar N(P',S') \left\{
     G_P(\Delta^2) \; \left[ \frac{\Delta_\mu^\perp}{2M} \gamma_5 \right]
   + G_A(\Delta^2) \; \left[ \gamma_\mu^\perp \gamma_5 \right]
\right\} N(P,S) .
\nonumber
\\
\label{A0mom}
\end{eqnarray}
Here the well-known sum rules \cite{Ji97a} have been used:
\begin{eqnarray}
\int_{-1}^{1}dx\, H(x, \xi, \Delta^2) &=& F_1(\Delta^2) ,
\nonumber
\\
\int_{-1}^{1}dx\, E(x, \xi, \Delta^2) &=& F_2(\Delta^2) ,
\nonumber
\\
\int_{-1}^{1}dx\, \widetilde H(x, \xi, \Delta^2) &=& G_A(\Delta^2) ,
\nonumber
\\
\int_{-1}^{1}dx\, \widetilde E(x, \xi, \Delta^2) &=& G_P(\Delta^2) ,
\label{JiSumRules}
\end{eqnarray}
where
$F_1(\Delta^2)$ and $F_2(\Delta^2)$ are the Dirac and Pauli form factors,
$G_P(\Delta^2)$ and $G_A(\Delta^2)$ are the pseudo-scalar
and axial-vector form factors; and, by definition, the magnetic form factor is
\begin{equation}
G_M(\Delta^2) = F_1(\Delta^2) + F_2(\Delta^2) .
\end{equation}
The results (\ref{V0mom}) and (\ref{A0mom}) imply that
$$
\int_{-1}^1 dx\ G_i(x,\xi,\Delta)=0,
\qquad
\int_{-1}^1 dx\ \widetilde G_i(x,\xi,\Delta)=0.
$$
These sum rules can be considered as a non-forward generalization
of Burkhardt-Cottingham \cite{Bur70} sum rules (\ref{WW0mom}).

%
%X^1 MOMENTS.
%

For the $x$-moments of the vector GPD we get:
\begin{eqnarray}
\label{V1_1mom}
\int_{-1}^{1}dx\, x\, G_1 (x, \xi, \Delta)
&=&
\frac12 \frac{\partial}{\partial \xi} \int_{-1}^{1}dx\, x\,  E(x, \xi, \Delta) ,
\\
\int_{-1}^{1}dx\, x\, G_2 (x, \xi, \Delta)
&=&
\frac12
\left[
     G_A(\Delta^2)
   - \int_{-1}^{1}dx\, x\, (H+E)(x, \xi, \Delta)
\right]  ,
\\
\int_{-1}^{1}dx\, x\, G_3 (x, \xi, \Delta) &=& 0  ,
\\
\int_{-1}^{1}dx\, x\, G_4 (x, \xi, \Delta) &=& 0  .
\label{V4_1mom}
\end{eqnarray}
Note that the $x$-moment of the GPD $G_2$ does not vanish in the forward limit and gives the
quark orbital momentum of the nucleon \cite{Pen00}:
$$
\lim_{\Delta\to 0}\int_{-1}^{1}dx\, x\, G_2 (x, \xi, \Delta)=-J_q+\frac12\ \Delta q=-L_q\, .
$$

For the $x$-moments of the axial-vector GPD we have:
\begin{eqnarray}
\label{A1_1mom}
\int_{-1}^{1}dx\, x\, \widetilde G_{1} (x, \xi, \Delta)
&=&
\frac12
   \left[
        F_2(\Delta^2)
      + \left( \xi \frac{\partial}{\partial \xi} - 1 \right)
           \int_{-1}^{1}dx\, x\, \widetilde E(x, \xi, \Delta)
   \right]  ,
\\
\int_{-1}^{1}dx\, x\, \widetilde G_{2} (x, \xi, \Delta)
&=&
\frac12
   \left[
        \xi^2 G_E(\Delta^2)
      - \frac{\Delta^2}{4M^2} F_2(\Delta^2)
      - \int_{-1}^{1} dx\, x\, \widetilde H (x, \xi, \Delta)
   \right]  ,
\\
\int_{-1}^{1}dx\, x\, \widetilde G_{3} (x, \xi, \Delta)
&=& \frac14 \, \xi \, G_E(\Delta^2)  ,
\\
\int_{-1}^{1}dx\, x\, \widetilde G_{4} (x, \xi, \Delta)
&=& \frac14 \, G_E(\Delta^2)  ,
\label{A4_1mom}
\end{eqnarray}
where
\begin{equation}
G_E(\Delta^2) = F_1(\Delta^2) + \frac{\Delta^2}{4M^2}\, F_2(\Delta^2)
\end{equation}
is the electric form factor. These moments have no gluon contribution due to (\ref{tw301mom})
and they have already been discussed in \cite{Pen00} and \cite{Kiv01a}.
Similarly to the previous case one can consider the relations
(\ref{V1_1mom})-(\ref{V4_1mom}) and (\ref{A1_1mom})-(\ref{A4_1mom})
as a non-forward generalization
of Efremov-Leader-Teryaev \cite{Efr97} sum rule (\ref{WW1mom}).

%
%X^2 MOMENTS.
%

We are passing now to a consideration of the $x^2$-moments. For these moments
the ``genuine twist-3'' (quark-gluon) contribution (\ref{tw32momperp}) does not vanish
and one can decompose the GPDs $G$ and $\widetilde G$
from (\ref{QVdec0}), (\ref{QAdec0}) into the WW and genuine twist-3 parts:
\begin{equation}
G_i^{(2)}(\xi, \Delta) = G_i^{(2) WW}(\xi, \Delta) + G_i^{(2) tw3}(\xi, \Delta) ,
\qquad
i= 1, 2, 3, 4.
\end{equation}
For the $x^2$-moments of the vector GPD ``kinematical'' part we get:
\begin{eqnarray}
\label{WWVG1}
\int_{-1}^{1} dx\, x^2 \, G_1^{WW}(x, \xi, \Delta) &=& \left(- \frac13 \right)
                     \left[
                  \xi F_2(\Delta^2)
                + \xi \left(
                     \xi \frac{\partial}{\partial \xi} - 1
                  \right) \widetilde E^{(1)}(\xi, \Delta)
            - \frac{\partial}{\partial \xi} E^{(2)}(\xi, \Delta)
             \right]   ,
\nonumber
\\
\\
\int_{-1}^{1} dx\, x^2 \, G_2^{WW}(x, \xi, \Delta)
                &=& \frac13 \left[
                                 \xi^2 G_M(\Delta^2)
                   + \frac{\Delta_{\perp}^2}{4M^2}
                        \left(
                           \xi \frac{\partial}{\partial \xi} - 1
                        \right)
                    \widetilde E^{(1)}(\xi, \Delta)
                \right.
\nonumber
\\
&& \qquad\qquad\qquad\qquad
              + \left.
                     \widetilde H^{(1)}(\xi,\Delta)
           - 2 ( H + E)^{(2)}(\xi,\Delta)
                \right] ,
\\
\int_{-1}^{1} dx\, x^2 \, G_3^{WW}(x, \xi, \Delta)
           &=&  \frac16
        \left[
                     \xi G_M(\Delta^2)
                   + \left(
                        1 - \frac{\Delta^2}{4M^2}
                     \right)
             \xi
             \left(
                        \xi \frac{\partial}{\partial \xi} - 1
                     \right)
             \widetilde E^{(1)}(\xi, \Delta)
                \right.
\nonumber
\\
&& \qquad\qquad\qquad\qquad
                \left.
                     {}
                   - \frac{\partial}{\partial \xi} \left( H + E \right)^{(2)}(\xi, \Delta)
                \right]   ,
\\
\int_{-1}^{1} dx\, x^2 \, G_4^{WW}(x, \xi, \Delta) &=& \left(- \frac16 \right)
                     \left(
            \xi \frac{\partial}{\partial \xi} - 1
             \right)
             \left[
                  \widetilde H^{(1)}(\xi, \Delta)
            + \frac{\Delta^2}{4M^2} \widetilde E^{(1)}(\xi, \Delta)
             \right]   ,
\label{WWVG4}
\end{eqnarray}
where we used shorthand notations for the Mellin moments of the twist-2 vector GPDs:
\begin{eqnarray}
H^{(m)}(\xi, \Delta) &=& \int^1_{-1} dx\, x^m H(x, \xi, \Delta) ,
\nonumber
\\
E^{(m)}(\xi, \Delta) &=& \int^1_{-1} dx\, x^m E(x, \xi, \Delta) ;
\nonumber
\end{eqnarray}
and analogous notations for the twist-2 pseudovector GPDs $\widetilde H$ and $\widetilde E$.

For the $x^2$-moments of the axial-vector GPD we get:
\begin{eqnarray}
\label{WWAG1}
\int_{-1}^{1} dx\, x^2 \, \widetilde G_1^{WW}(x, \xi, \Delta)
&=& \left( - \frac13 \right)
    \left( \xi \frac{\partial}{\partial \xi} - 2 \right)
    \left[
       E^{(1)}(\xi, \Delta) - \widetilde E^{(2)}(\xi, \Delta)
    \right] ,
\\
\int_{-1}^{1} dx\, x^2 \, \widetilde G_2^{WW}(x, \xi, \Delta)
&=& \frac13
    \Biggl[
       \xi^2 \left(
                G_A(\Delta^2) +  \left( H + E \right)^{(1)}(\xi, \Delta)
             \right)
\nonumber
\\
&&  \qquad
       {} + \frac{\Delta_{\perp}^2}{4M^2}
            \left(
               \xi \frac{\partial}{\partial \xi} - 2
            \right)
        E^{(1)}(\xi, \Delta)
      - 2 \widetilde H^{(2)}(\xi, \Delta)
    \Biggr]   ,
\\
\int_{-1}^{1} dx\, x^2 \, \widetilde G_3^{WW}(x, \xi, \Delta)
&=& \frac16
    \Biggl[
       \xi \biggl(
              G_A(\Delta^2) + ( H + E )^{(1)}(\xi, \Delta)
           \biggr)
\nonumber
\\
&+&
         \left(
               1 - \frac{\Delta^2}{4M^2}
         \right)
         \xi
         \left(
            \xi \frac{\partial}{\partial \xi} - 2
         \right)
         E^{(1)}(\xi, \Delta)
       - \frac{\partial}{\partial \xi} \widetilde H^{(2)}(\xi, \Delta)
    \Biggr]  ,
\\
\int_{-1}^{1} dx\, x^2 \, \widetilde G_4^{WW}(x, \xi, \Delta)
&=& \left( - \frac16 \right)
    \left( \xi \frac{\partial}{\partial \xi} - 2 \right)
    \left[
         H^{(1)}(\xi, \Delta)
       + \frac{\Delta^2}{4M^2} E^{(1)}(\xi, \Delta)
    \right]   .
\label{WWAG4}
\end{eqnarray}

Let us consider the quark-gluon contributions to the moments of ${\cal F}_\perp$
and $\widetilde {\cal F}_\perp$.
We have already shown that their $x^0$- and $x^1$-moments vanish (\ref{tw301mom}),
with the $x^2$-moment being given by (\ref{tw32momperp}) in the vector case.
Substituting (\ref{Vnucl}) into (\ref{tw32momperp}) and decomposing the result
according to (\ref{QVdec0}) we get the following expressions for the $x^2$-moments
of the genuine twist-3 vector GPD at $\Delta^2=0$:
\begin{eqnarray}
\label{tw3VG1}
\int_{-1}^{1} dx\, x^2 \, G_1^{tw3}(x, \xi)
&=&
{} 0,
\\
\int_{-1}^{1} dx\, x^2 \, G_2^{tw3}(x, \xi)
&=&
{} - \frac23\, (1-\xi^2) \,  d^{(2)} ,
\\
\int_{-1}^{1} dx\, x^2 \, G_3^{tw3}(x, \xi)
&=&
\frac13\, \xi \, d^{(2)} ,
\\
\int_{-1}^{1} dx\, x^2 \, G_4^{tw3}(x, \xi)
&=&
\frac13 \, d^{(2)}.
\label{tw3VG4}
\end{eqnarray}
Here $d^{(2)}$ is given by Eq.~(\ref{d2}).
For the axial-vector GPD we get at $\Delta^2=0$:
\begin{eqnarray}
\label{tw3AG1}
\int_{-1}^{1} dx\, x^2 \, \widetilde G_1^{tw3}(x, \xi)
&=& 0,
\\
\int_{-1}^{1} dx\, x^2 \, \widetilde G_2^{tw3}(x, \xi)
&=&
\frac23 \, (1-\xi^2) \, d^{(2)},
\\
\int_{-1}^{1} dx\, x^2 \, \widetilde G_3^{tw3}(x, \xi)
&=&
{} - \frac13\, \xi \, d^{(2)}  ,
\\
\int_{-1}^{1} dx\, x^2 \, \widetilde G_4^{tw3}(x, \xi)
&=&
{} - \frac13 \, d^{(2)}.
\label{tw3AG4}
\end{eqnarray}

The expressions (\ref{tw3VG1})-(\ref{tw3VG4})  and (\ref{tw3AG1})-(\ref{tw3AG4})
for the the genuine twist-3 contributions should be compared to the expressions
(\ref{WWVG1})-(\ref{WWVG4}) and (\ref{WWAG1})-(\ref{WWAG4}) of the WW part for the quark contributions
correspondingly. As we saw above, in the instanton vacuum the constant $d^{(2)}$ is strongly suppressed
by the packing fraction of instantons \cite{Bal97}. Here we expressed the $x^2$-moments of the genuine
twist-3 GPDs in terms of the constant $d^{(2)}$, see expressions (\ref{tw3VG1})-(\ref{tw3VG4})  and
(\ref{tw3AG1})-(\ref{tw3AG4}). We conclude that the twist-3 vector and axial-vector GPDs
are parametrically suppressed relative to the WW ones due to the packing fraction
of the instanton vacuum:
\begin{equation}
\frac{G_i^{(2) tw3}}{G_i^{(2) WW}}
\sim \pi^2\left( \frac{\rho^2}{R^2} \right)^2
     \ln \left( \frac{\rho^2}{R^2} \right)
\ll  1   .
\end{equation}
This indicates that in the QCD vacuum the ``genuine twist-3'' (quark-gluon) contributions
turn out to be small relative to the purely quark ones.

\section*{Acknowledgments}
We acknowledge useful discussions with N.~Kivel, A.~Mukherjee and M.~Vanderhaeghen.
We thank Vadim Guzey for careful reading of the manuscript.
This work is supported
by the Sofja Kovalevskaja Programme of the Alexander von Humboldt
Foundation, the Federal Ministry of Education and Research and the
Programme for Investment in the Future of German Government.

\end{document}